# Ti$_2$NiCu Based Composite Nanotweezers with a Shape Memory Effect and its Use for DNA Bunches 3D Manipulation

A.P. Orlov[1,2,a)], A.V. Frolov[1], A.M. Smolovich[1], P.V. Lega[1,b)], P.V. Chung[1],
A.V. Irzhak[1], N.A. Barinov[3], D.V. Klinov[3], V.S. Vlasenko[4] and V.V. Koledov[1]

[1]*Kotelnikov Institute of Radioengineering and Electronics of RAS, Mokhovaya 11-7, Moscow, 125009, Russia,*
[2]*Institute of Nanotechnology of Microelectronics of RAS, Nagatinskaya 16A/11, Moscow, 115487, Russia.*
[3]*Federal Research and Clinical Center of Physical-Chemical Medicine, 119435, Moscow, Russia*
[4]*OPTEC LLC, Denisovsky lane, 26, Moscow, 105005, Russia*

[a)] Corresponding author: andreyorlov@mail.ru
[b)] lega_peter@list.ru

**Abstract.** The DNA molecules were controllable deposited on graphene and thin graphite films and visualized using AFM. The mechanical micro- and nanotools, such as nanotweezers with shape memory effect controlled by heating were designed and tested. A technique for fabricating a structure with the inclusion of suspended DNA threads and manipulating those using composite nanotweezers with shape memory effect was suggested.

## INTRODUCTION

The modern paradigm of the life science is thoroughly based on the DNA role in conservation and expression of genetic information since it was extracted by Miescher in 1869 and its double strands helix structure was identified by Watson, Crick and Franklin in 1953. For long time after these discoveries the progress of the study and processing of DNA in biology is based mainly on the multiple molecules approach by chemical methods. An individual DNA molecules nanomanipulation has started in 1992, when AFM was used for dissection of single DNA strands by an AFM tip on a solid surface [1]. Since then more and more complicated 2D manipulations procedures with DNA by AFM were successfully implemented [2, 3]. Even some 3D picking up and positioning were realized by AFM with individual DNA samples [4].

Modern nanotechnology exploits the outstanding molecular selectivity properties of DNA and other nucleic acids to create self-assembling branched DNA complexes with useful properties. DNA is thus used as a structural material rather than as a carrier of biological information. This has led to the creation of two-dimensional periodic lattices (both tile-based and using the DNA origami method) and three-dimensional structures in the shapes of polyhedra. Nanomechanical devices and algorithmic self-assembly have also been demonstrated and these DNA structures have been used to template the arrangement of other molecules such as gold nanoparticles and streptavidin proteins. DNA is used by researchers as a molecular tool to explore physical laws and theories, such as the ergodic theorem and the theory of elasticity. The unique material properties of DNA have made it an attractive molecule for material scientists and engineers interested in micro- and nano-fabrication. Among notable advances in this field are DNA origami and DNA-based hybrid materials, which are promising for the next step nanoelectronics and nanophotonics [5-8].

Thus the individual nanomanipulation of DNA molecules has led to improvement not only AFM, but other micromanipulation systems optical systems, acoustics, automotive, etc. which in principal, could provide 3D nanoprocessing and nanomanipulation. Among them are electrostatic and magnetic nanotweezers [9, 10].

The purpose of the present paper is to consider in detail the new concept of shape memory nanotweezers [11-13] applied specifically for macromolecules bunches manipulation. The physical principals and nanotechnological approaches to engineering and control of the processes of DNA manipulation will be outlined together with preliminary experiments on the co-processing of DNA on graphene substrate.

## 1. DESIGN THERMALLY-CONTROLLED NANOTWEEZERS

The reversible deformation of the composite on thermal cycling is not an intrinsic property of intermetallics undergoing thermoelastic martensitic transition [11]. Fig. 1 illustrates the principles of the controlled bending strain of a bilayer composite, including an elastic metallic layer and the ribbon of the alloy, such as $Ti_2NiCu$, exhibiting SME, which is initially stretched in the martensitic condition [12, 13]. The martensitic transition in $T_2NiCu$ takes place in the range of 42-52°C. The processes of connecting the elastic and active layers should be done below martensitic transition temperature. In this case the resulting composite demonstrates controlled reversible strain on thermally cycling it through the austenite- martensite transition temperature range (see Fig. 2.) [12-16]. Melt spun ribbons of the alloy $T_2NiCu$, are chosen as the basic functional material because they possess of outstanding SME, suitable temperature range of transition, easy and economy of production [17-20].

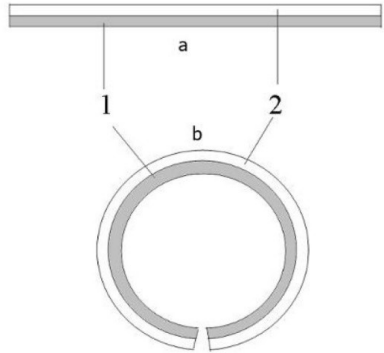

**FIGURE. 1**. Schematic diagram of a bimorph composite, capable of reversible bending deformation: (1) layer with SME in the (a) low temperature martensite and (b) high temperature austenite state; (2) elastic metal layer [12-13].

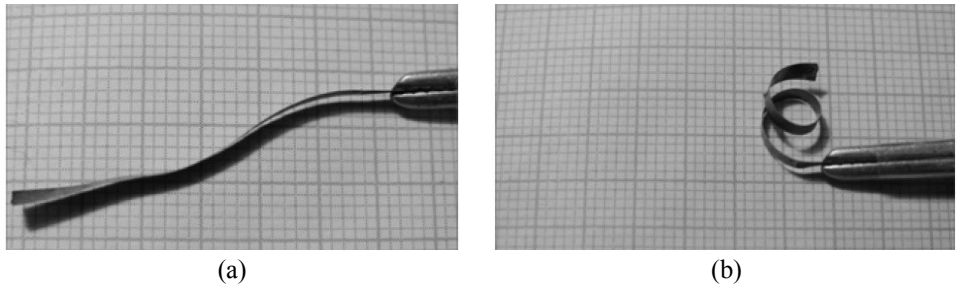

(a)          (b)

**FIGURE 2.** Reversible bending deformation of $Ti_2NiCu/Ni$ composite manufactured by galvanic deposition of nickel onto a rapidly quenched ribbon of the alloy Ti2NiCu with SME: (a) composite in the martensitic state at room temperature; (b) SMA layer is in the austenitic state after heating above 56°C. The composite demonstrates long term reversible deformation (of up to thousands of thermal cycles [13]).

Modern FIB systems allow manufacturing of the microactuators from 100 μm to 1 μm in length, with the active layer thickness in the range of 100-1000 nm. The relative strain that the composite undergoes (~1%) is controlled by heating and the magnitude of strain is similar to that of the alloy in the bulk form. The preparation of $Ti_2NiCu/Pt$ composite microtweezers is done by FIB CVD (for more details see [21-28]). The main stages of FIB etching of nanotweezers from the $Ti_2NiCu$ ribbon witch amorphous layer are illustrated in Fig 3.

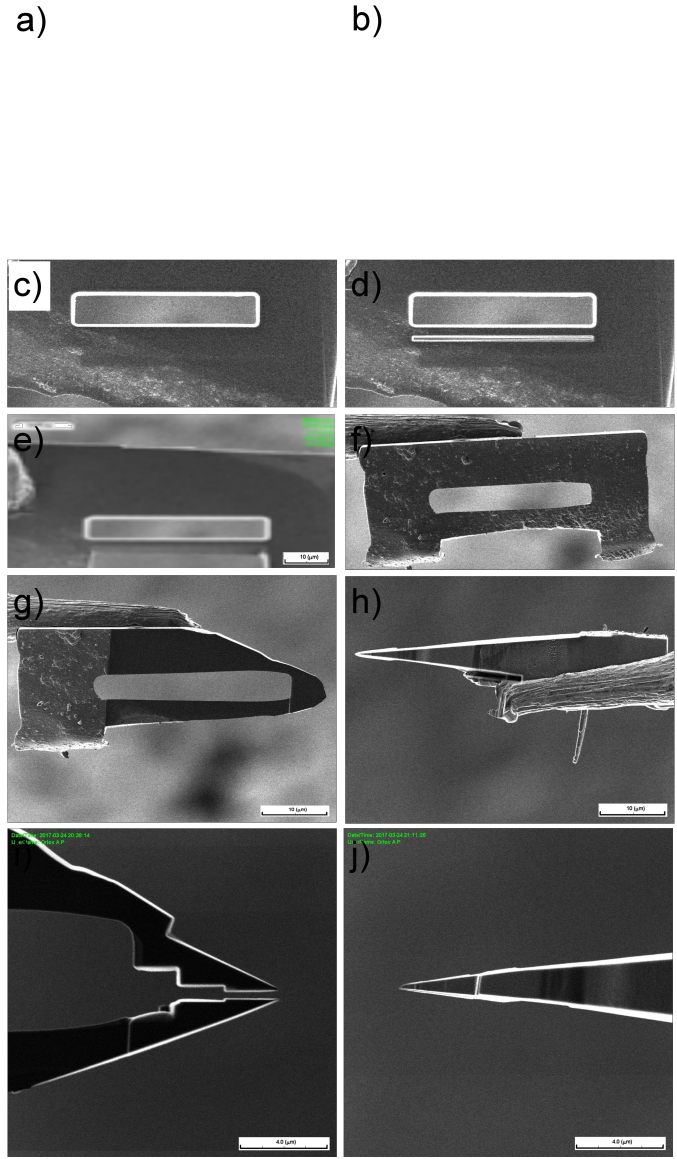

**FIGURE 3**. The main stages of FIB etching of nanotweezers (a) the blank tapes for tweezers etched on a tape made of SME - Ti2NiCu material; (b) tweezers blank transferred on the needle with the mini heater; (c) etching of the first window that determines the shape of tweezers; (d) etching of the second window that forms a movable part; (e) spraying of the amorphous layer on the working beam; (f) design procurement of tweezers blank on the needle; (g,h) etching and cleaning the edges of the sharpened working tip of the nanotweezers, (i,j) forming and cleaning a working gap on the finished nanotweezers.

For the tasks of DNA manipulations, authors have been designed nanotweezers attached to a tungsten needle installed in manipulators Kleindiek in scanning electron microscope (SEM) CrossBeam Neon 40 EsB (Carl Zeiss). Fig. 4 shows gripping part of the nanotweezers from $Ti_2NiCu$ alloy with working gap of ~170 nm in open (martensitic) state and in close (austenitic) state.

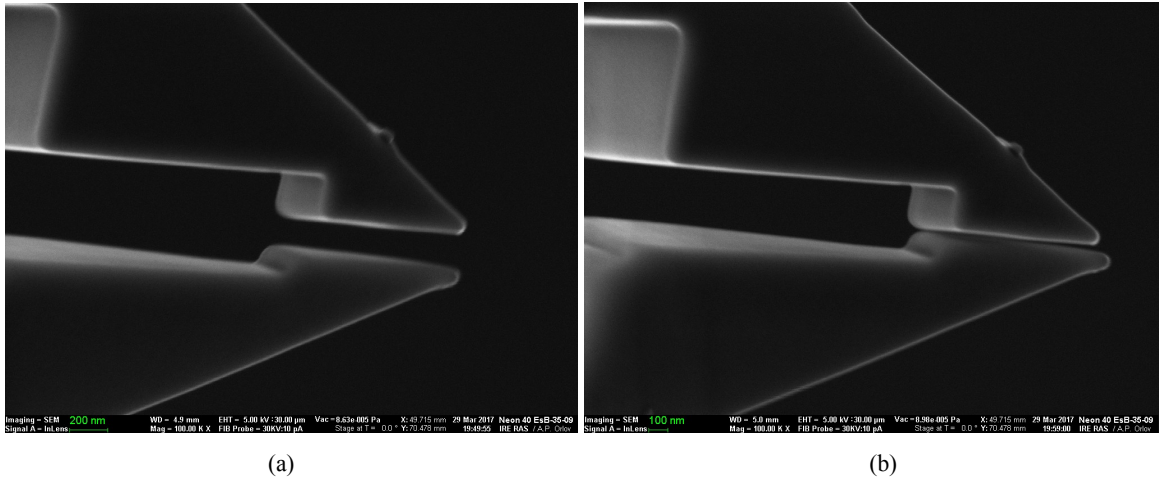

**FIGURE 4**. SEM image composite nanotweezers from Ti2NiCu alloy after multiple testing (a) in open state and (b) in close state.

## 2. FABRICATING OF SUSPENDED DNA BUNCHES AND MANIPULATING THOSE

In experiments on DNA manipulation by nanotweezers, a membrane of nitrocellulose was first used as a carrier for DNA molecules. The membrane thickness was about 50 nm. The membrane was placed on a copper mesh for a transmission microscope with square cells with a side of 50 μm, or alternatively the membrane was placed on a gap between two silicon wafers. The amorphous carbon layer with a thickness of the order of 10 nm was sputtered upon the membrane surface for its strengthening. Then, using a focused ion beam (FIB) SMI3050, rectangular narrow (0.1-2 μm) cuts 10-20 μm in length were etched in the membrane (Fig. 5a). Also, we tried to use the other shapes of the FIB cuts instead of rectangular ones (Fig. 5b).

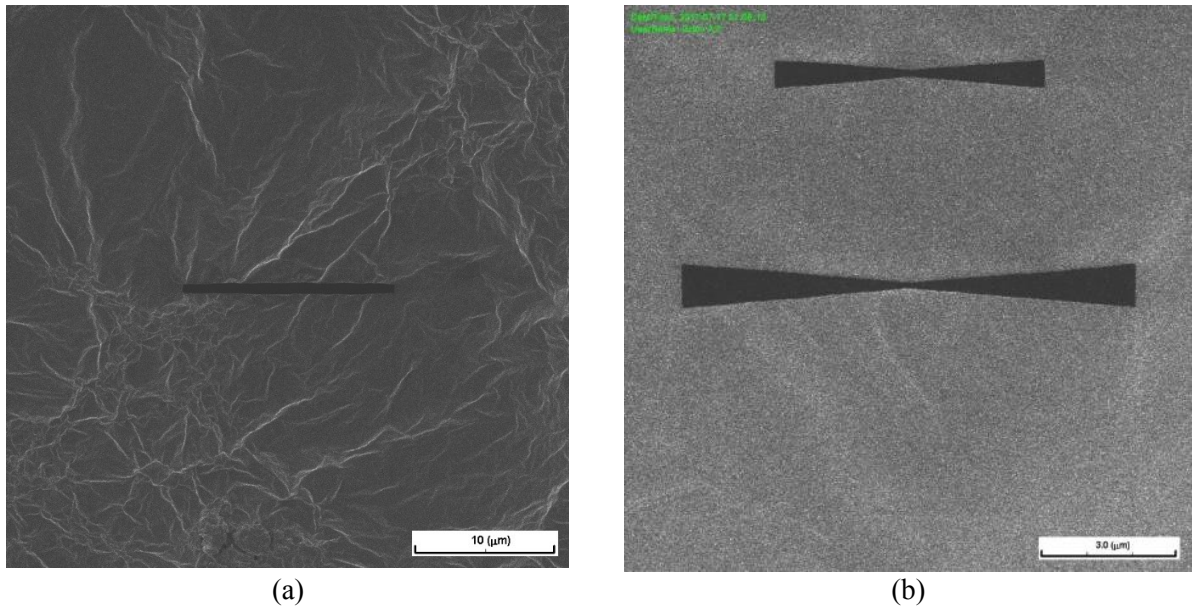

**FIGURE 5**. Membrane with FIB cuts of different shapes: rectangular (a) and special shape (b).

The other option was using a graphene-thin graphite layer for the preparation of suspended DNA samples. Thin graphite layers were also placed on a copper mesh or on a gap between two thin pieces of cover glass. The narrow cuts etched by FIB were also performed on the thin graphite layers. In general, the technology of applying DNA molecules on the graphene-thin graphite layers or the nitrocellulose membrane was similar to the procedure for deposition DNA molecules to graphite [29-32]. However, some difference described below took place. Our experience of deposition and visualization of DNA molecules on graphene [33] was also used.

Solution of graphite modifier (GM) (($CH_2$)$_n$($NCH_2CO$)$_m$–$NH_2$ from Nanotyuning, Russia) with a concentration of 0.1% and a volume of 100 μL is deposited on the graphene surface. After 1-min-long exposure, the GM is removed using nitrogen jet and the substrate is dried. Molecules of duplex DNA from Escherichia virus Lambda with a concentration of 1 μg/mL are deposited from the solution containing 10 mM Tris-HCl (pH 7.6) and 1 mM EDTA on the surface of the modifier for 1 min, and, then, the drop with dissolved DNA is removed using the nitrogen jet. The monolayer of the modifier represents lamellar structures that are epitaxially crystallized by intermolecular H bonds on the graphene surface. The deposition of the modifier is needed for attachment of unfolded DNA molecules to the surface, since DNA molecules weakly interact with pure graphene and exhibit twisting, folding, and shifting upon passage of the droplet meniscus. When the droplet is dried rather than removed using the nitrogen jet, the impurities contained in the solution at mass fractions of greater than $10^{-6}$ are precipitated as a rough layer with a thickness of about 1 nm that is comparable with the DNA thickness. This circumstance impedes the DNA identification using AFM.

The operation of a drop with GM or DNA solution removing by nitrogen jet was not possible for the suspended on the coper mesh nitrocellulose membrane or graphene as underlying material by the reason or material destruction In this case a drop of DNA was removed by drying without blowing, with a piece of filter paper.

We employ AFM to monitor the DNA deposition [34]. The landscapes of samples are measured in the semicontact resonance regime on an NT-MDT Integra Prima setup using NOVA 1.1 software. High-resolution supersharp silicon cantilevers from Nanotyuning are used for the measurements. The resonance frequencies of the cantilevers range from 190 to 325 kHz, the radius of curvature of the tip is less than 2 nm, and the apex angle is less than 22º. The amplitude of the free oscillations of the cantilever in air ranges from 1 to 10 nm. The automatically maintained amplitude of the cantilever oscillations in the vicinity of the surface (SetPoint parameter) is fixed at a level of 60–70% of the amplitude of the cantilever oscillations in air.

The NOVA 1.1 software from NT-MDT is used for signal processing, digitization, and imaging. The experimental results are represented as 2D images in which light and dark surface fragments correspond to hills and wells, respectively (Figs. 6). The NOVA Image Analysis 2.0 software is used for data processing.

Figure 6 presents the AFM images of DNA molecules on the graphene surface. It is seen that the proposed method makes it possible to obtain relatively uniform distribution of DNA on the graphene surface (Fig. 6a). DNA molecules form relatively strong bonds with modified graphene, so that even small modifier-free fragments are covered with DNA molecules. The measurements of heights in such fragments are used to estimate the thickness of the modifier layer (about 0.7 nm).

Practical results show that the modifier must be deposited on graphene immediately after splitting, since relatively long exposure of the graphene surface to air leads to contamination, so that the modifier cannot be deposited after short exposures. The graphite surface with the modifier is more stable in air: DNA was successfully deposited on the graphene sample that was stored under forvacuum conditions over three months after deposition of the modifier (Fig. 6b). For such a sample, the DNA molecules are stretched along the direction of the gas jet owing to relatively weak interaction with the modifier. For better adhesion of DNA to modifier, the droplet with dissolved DNA is kept on the sample over a longer time interval (2 – 3 min). However, in this case, the deposition of DNA is accompanied by precipitation of impurities: balls with a diameter of several nanometers are observed on the graphene surface (Fig. 6b).

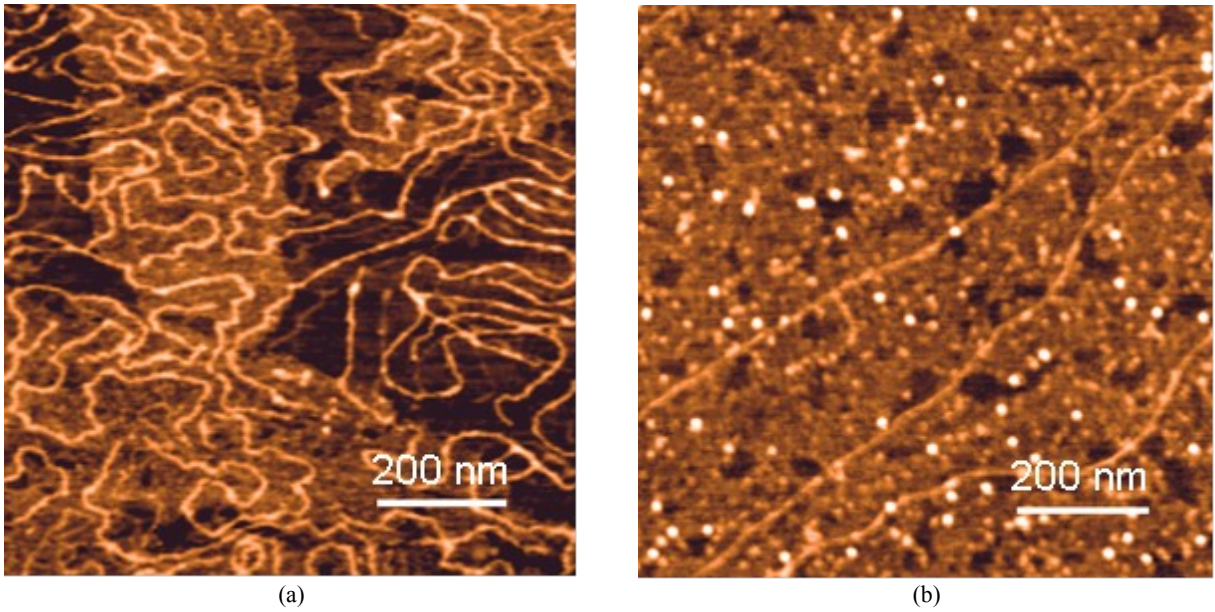

**FIGURE 6.** AFM images of DNA molecules on (a) graphene surface with the modifier layer, (b) graphene samples stored under forvacuuum conditions after modifier deposition.

AFM image of DNA molecules on a membrane applied by drying without blowing is shown in Fig. 7. We can see in the figure that the DNA is mixed with impurities. We suppose that this is consequence of using DNA drop removing by drying with a piece of filter paper without blowing. The other disadvantage of this drying procedure is the obtaining of DNA bunches instead of solitary DNA molecules. However, manipulation with DNA bunches is easier due to their bigger thickness and better visibility by scanning electron microscope (Fig. 8). So, we began manipulation with DNA bunches as the first step. The two phases of capturing DNA bunch by composite nanotweezers is shown on Fig. 9a and Fig. 9b.

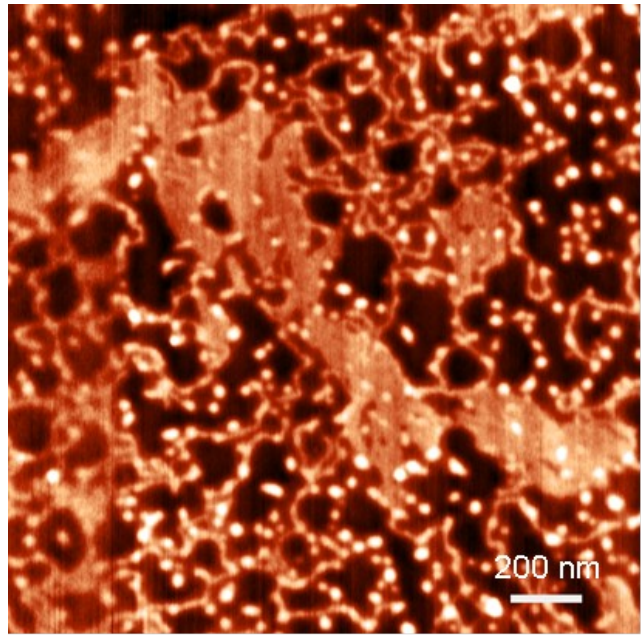

**FIGURE 7.** High-resolution atomic force microscopy of DNA molecules on a membrane applied by drying without blowing (in the figure, the DNA is mixed with impurities). Colors in topography images represent the height variation $\Delta h = 10$ nm.

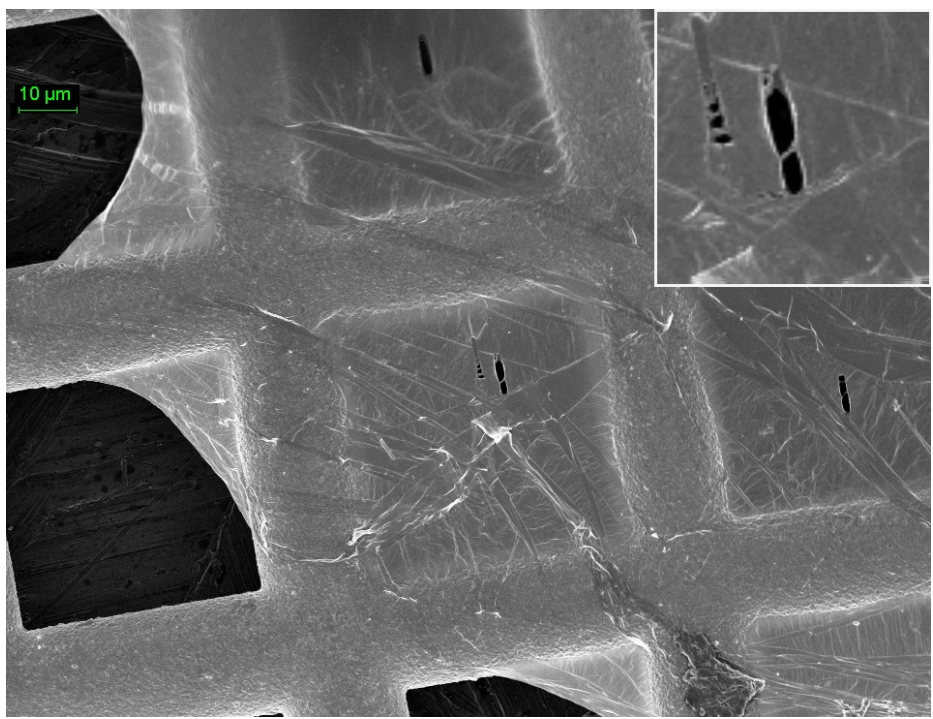

**FIGURE 8.** SEM image of nitrocellulose membrane on copper mesh with FIB cuts after DNA deposition. Inset: the DNA bunch planted in cut.

Using nanotweezers it is possible to grab DNA tightly, squeezing it on both sides. Moving tweezers we can not only stretch the DNA, but also tear the molecule off the edge of the membrane, as shown in Fig. 9. Thus the membrane is fixed between two electrical contacts, where one contact is a conductive membrane, the other one is conductive nanotweezers. Such incorporation of DNA into the electrical circuit is of scientific interest for studying the electrical conductivity of DNA molecules. This technique can also be used for the study of nanotube and nanowires.

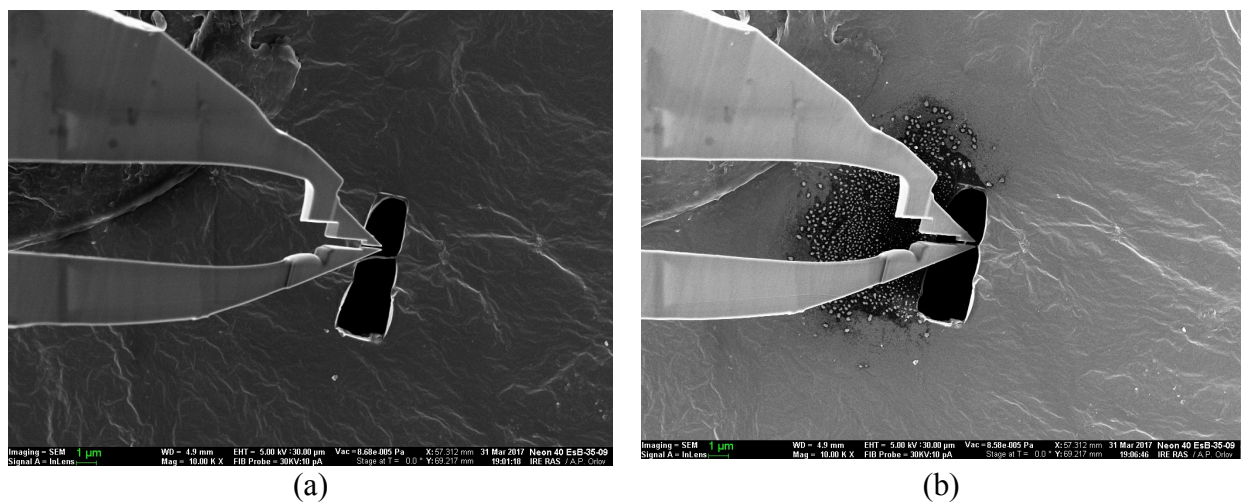

(a)      (b)

**FIGURE 9**. Two phases ((a) and (b)) of capturing DNA bunch by composite nanotweezers.

## CONCLUSION

The suspended DNA bunches were fabricated by deposition of DNA molecules on nitrocellulose membranes or thin graphite films with FIB-made cuts. Manipulating DNA bunches by composite nanotweezers with shape memory effect was demonstrated. This technique could be suitable for DNA studying in transmission electron microscope without using the substrate and for studying its electrotransport properties under tension directly in SEM.

## ACKNOWLEDGMENTS

The work was supported by RSF – grant No 17-19-01748.